 \documentclass[final,1p,times,twocolumn]{elsarticle}

\usepackage{amssymb}
\usepackage{amsthm}
\usepackage{color}
\usepackage{ulem}
\usepackage{listings}
\usepackage{verbatim}
\usepackage{amsmath}

\numberwithin{equation}{section}

\journal{Applied Mathematics and Computation}

\begin{document}

\begin{frontmatter}



\title{An Efficient Algorithm of Logarithmic Transformation to Hirota Bilinear Form of KdV-type Bilinear Equation }


\author{Yichao Ye}

\author{Lihong Wang}

\author{Zhaowei Chang}

\author{Jingsong He\corref{cor1}}
\ead{hejingsong@nbu.edu.cn}
\cortext[cor1]{Corresponding author}

\address{Department of Mathematics, Ningbo University, Ningbo Zhejiang 315211, P.R. China}

\begin{abstract}

In this paper, an efficient algorithm of logarithmic transformation to Hirota bilinear form
of the KdV-type bilinear equation is established. In the algorithm, some properties of Hirota
operator and logarithmic transformation are successfully applied, which helps to prove that
the linear terms of the nonlinear partial differential equation play a crucial role in finding
the Hirota bilinear form. Experimented with various integro-differential equations, our algorithm
is proven to be more efficient than the algorithm referred by Zhou, Fu, and Li in getting the
Hirota bilinear form, especially in achieving the coefficient of the logarithmic transformation.

\end{abstract}

\begin{keyword}
Hirota bilinear form, KdV-type bilinear equation,
 D-operator, Logarithmic transformation, Symbolic computation.


\end{keyword}

\end{frontmatter}


\section{Introduction}

In 1971, Hirota developed a direct method,  Hirota bilinear method,
to construct the exact multi-soliton solution of integrable
nonlinear partial differential equation (NPDE) [1-3]. Once the Hirota bilinear
form(HBF) of a NPDE is given, there are several ways to solve it. The solutions of NPDE can be constructed from HBF by the typical perturbation expansion method \cite{1} and others [4,5].
Therefore, the key step of the Hirota direct method is to transform the NPDE into the Hirota bilinear form (HBF).
Hietarinta designed a program on searching for integrable
bilinear equations such as the KdV-type \cite{2}, mKdV-type \cite{3},
SG-type \cite{4}, NLS-type \cite{5}  equations. In 1992, Hereman and
Zhuang \cite{6} gave a summary on types of bilinear equations.
In recent years, some algorithms for generating bilinear form of NPDE are
described and Maple packages Bilinearization and HBFTrans are
established by Zhou, Fu, Li \cite{7,8} and Yang, Ruan \cite{9},
respectively. The package Bilinearization can construct the HBF of
many NPDEs by solving a system of over-determined algebraic
equations with respect to the combinatorial coefficients. However,
the general ansatz of the bilinear form of NPDE in Ref. \cite{7,8}, which relies on the \textbf{WTC} method \cite{10} and the \textbf{HBM} method \cite{11}, is so complicated that the efficiency of performance is relatively low. Without depending on these two methods, our method can obtain the HBF of the KdV-type equations within shorter time. In this paper, some properties of Hirota-operator are taken advantage of, which brings forward to a
more efficient algorithm for finding the HBF of KdV-type equations in NPDE. We take a series of classic KdV-type equations for instance, to
demonstrate the validity of our algorithm. Furthermore, the implementation of the
algorithm in Maple is applied to automate the tedious computation for the
construction of the HBF of KdV-type equations in NPDE.

\section{Hirota Bilinear Method}

Now, let us  briefly review the Hirota derivatives. In 1971, Hirota
developed the Hirota derivative, which is also called D-operator \cite{1}. For $1+1$ dimensions, the D-operator is defined by
\begin{equation}
 D_{x}^{n}D_{t}^{m}f \cdot
g=(\partial_{x}-\partial_{x'})^{n}(\partial_{t}-\partial_{t'})^{m}f(x,t)g(x',t')
\big|_{x'=x,t'=t} , \qquad m,n=0,1,2,3,\cdots .
\end{equation}
where $f(x,t)$ and $g(x,t)$ are differentiable functions of $x$ and
$t$, respectively.

From the definition, there are some properties \cite{1} of the D-operator,
\begin{align}
&D_{x}^{n}D_{t}^{m}f\cdot g=D_{t}^{m}D_{x}^{n}f\cdot
g=D_{x}^{n-1}D_{t}^{m}D_{x}f\cdot g\\
&D_{x}^{n}D_{t}^{m}f\cdot 1= \partial_x^n\partial_t^m f ,\quad
\text{where } \partial_{x}^n \equiv \partial^n/\partial{x^n}.\\
&D_{x}^{n}D_{t}^{m}f\cdot g=(-1)^{m+n}D_{x}^{n}D_{t}^{m}g\cdot f\label{property1}\\
&D_{x}^{n}D_{t}^{m}f\cdot f=0, \quad\text{if } m+n\text{ is odd}\label{property2};
\end{align}
furthermore, for arbitrary independent variable $x$, a recursive definition of D-operator is,
\begin{equation}
D_{x}^{n}f\cdot g =
\begin{cases}
fg,                & \text{for  $\quad n=0$} \\
D_{x}^{n-1}f_x\cdot g - D_{x}^{n-1}f\cdot g_x, & \text{for  $\quad n>0$} .\\
\end{cases}
\end{equation}

As an example, the Korteweg de Viries (KdV) equation
\begin{equation}
u_{xxx}+6uu_x+u_t=0\
\end{equation}
 where $u=u(x,t)$, can be
transformed through the dependent variable transformation,
\begin{equation}
u=2(\ln{f})_{xx}
\end{equation}
 into
\begin{equation}
\frac{\partial}{\partial x}\left(\frac{f_{xt}f - f_xf_t +
f_{xxxx}f-4f_{xxx}f_x + 3f^2_{xx}}{f^2}\right)=0,
\end{equation}
from which the bilinear equation below is obtained
\begin{equation}\label{bieqn}
f_{xt}f - f_xf_t + f_{xxxx}f-4f_{xxx}f_x + 3f^2_{xx}=cf^2,
\end{equation}
 where c is a constant of integration. Equation
(\ref{bieqn}), with c = 0, may also be written concisely in terms of
D-operators as

\begin{equation}
(D_{x}^{4}+D_{x}D_{t})f\cdot f=0.
\end{equation}
Thus,  the Hirota bilinear form of the KdV equation is obtained.

\section{Principles Of Our Algorithm}

\subsection{The Properties of Hirota Derivatives}

Introducing vector notation
\begin{equation}
\overrightarrow{D}=(D_t,D_x,D_y,\cdots),
\end{equation}
 consider the KdV-type bilinear equation \cite{1}
\begin{equation}
P(D_t,D_x,D_y,\cdots) f\cdot f = 0,
\end{equation}
where $P$ is a general polynomial in $D_t,D_x,D_y,\cdots$.

According to (\ref{property1}) and (\ref{property2}), it follows that
\begin{equation}\label{P_even}
P(-\overrightarrow{D})=P(\overrightarrow{D}).
\end{equation}
Now assume that the degree of every term in $P$ is even.

When $n$ is even, it's true that
\begin{equation}
\begin{split}
D_{x}^{n}f\cdot f
&=
\left(\sum_{i=0}^{n}C_n^i\partial_x^{n-i}(-\partial_{x'}^{i})\right)f(x)f(x')
\big|_{x'=x}\\
&=
\sum_{i=0}^{n}(-1)^iC_n^i\left(\partial_x^{n-i}f(x)\right)\left(\partial_{x}^{i}f(x)\right) \\
&=
2\sum_{i=0}^{{n}/{2}-1}(-1)^iC_n^i\left(\partial_x^{n-i}f(x)\right)\left(\partial_{x}^{i}f(x)\right) + (-1)^{{n}/{2}}C_n^{n/2}\left(\partial_x^{n/2}f(x)\right)^2 \\
&= 2\left(\partial_x^{n}f\right)f+ H,
\end{split}
\end{equation}
where $H=H(f_x,f_t,\cdots)$ is a polynomial in $f_x,f_t,\cdots$, WITHOUT $f$, and
$H=D_{x}^{n}f\cdot f - 2\left(\partial_x^{n}f\right)f$.
This simple observation implies the following crucial formula
\begin{equation}\label{bPD1bpartial}
P(D_t,D_x,D_y,\cdots) f\cdot f = 2
\left(P(\partial_t,\partial_x,\partial_y,\cdots) f\right)f + \widetilde{H}=2\left(P(\overrightarrow{\partial})f\right)f+
\widetilde{H},
\end{equation}
where $\widetilde{H}$ is a polynomial in $f_x,f_t,\cdots $, WITHOUT $f$, and
$P(\overrightarrow{\partial})=P(\partial_t,\partial_x,\partial_y,\cdots)$.
Some simple examples are listed as follows to illustrate equation (\ref{bPD1bpartial})
\begin{align}
&D_{x}^{2}f\cdot f=2({f_{xx}f}-f_{x}^2),\;\; \widetilde{H}=-2f_{x}^2 , \\
&D_{x}^4f\cdot f=2({f_{4x}f}-4f_{3x}f_{x}+3f_{xx}^2),\;\;\widetilde{H}=-8f_{3x}f_{x}+6f_{xx}^2,\\
&D_{x}D_{t}f\cdot f= 2({f_{tx}f}-f_{t}f_{x}),\; \;\widetilde{H}=-2f_{t}f_{x},
\end{align}
where $f_{kx}\equiv \partial_x^k f\equiv\frac{\partial^k f}{\partial x^k}$,  $k\in \mathbb{N}$.
Furthermore, using equation (\ref{bPD1bpartial}), it is easy to find $P(\overrightarrow{D})f \cdot f$
is a quadratic polynomial in $f,f_x,f_t,\cdots$, and then to obtain
\begin{equation}\label{HBF/f^2}
\dfrac{P(\overrightarrow{D})f \cdot
 f}{f^2}=\dfrac{P_1}{f}+\dfrac{P_2}{f^2}.
\end{equation}
Here $P_1$ and $P_2$ are polynomials in $f_x,f_t,\cdots$, WITHOUT $f$.
$P_1=2P(\overrightarrow{\partial})f$, is a linear polynomial; $P_2=\widetilde{H}$ is a
quadratic polynomial.

\subsection{Logarithmic Transformation}

Consider the derivation formula,
\begin{equation}\label{diff_formula}
\left( \dfrac{f}{g} \right)_x =
\dfrac{f_xg-fg_x}{g^2}=\dfrac{f_x}{g}-\dfrac{fg_x}{g^2}.
\end{equation}
Furthermore, for arbitrary independent variable $t$, it's easy to show that
\begin{equation}
\left( \dfrac{f_x}{f} \right)_{kt} =\dfrac{\overline{P}_1}{f}+\dfrac{\overline{P}_2}{f^2}
+ \cdots + \dfrac{\overline{P}_{k+1}}{f^{k+1}},
\end{equation}
where $k\in \mathbb{N}$ and $\overline{P}_i(i=1,2,\cdots,k+1)$ is a homogeneous polynomial of degree $i$
in $f_x,f_t,\cdots$, WITHOUT $f$, and
\begin{equation}
\overline{P}_1=\partial_t^kf_x, \; \overline{P}_{k+1}=(-1)^{k}k!f_{x}f_{t}^{k}.
\end{equation}
are obtained by using equation (\ref{diff_formula}) repeatedly.

Furthermore, when $m+n\geq 1$, it shows that
\begin{equation}\label{lnf_diff}
\partial_x^n\partial_t^m\left(\ln{f}\right)=\frac{\partial_x^n\partial_t^mf}{f}+\cdots+(-1)^{m+n-1}(m+n-1)!\frac{f_x^nf_t^m}{f^{m+n}}
\end{equation}

Still, by using $P$ as the denotation of a general polynomial and letting the lowest degree of the
terms in $P$ be equal to or greater than $1$, we have
\begin{equation}\label{lnf_diff2}
P(\overrightarrow{\partial})(\ln{f})=\frac{\widetilde{P}_1}{f}+\frac{\widetilde{P}_2}{f^2}+\cdots+\frac{\widetilde{P}_r}{f^r},
\end{equation}
where $r=\mathrm{degree}(P)$  and $\widetilde{P}_1=P(\overrightarrow{\partial})f$. Here, $\widetilde{P}_i(i=1,2,\cdots,r)$  also denotes a homogeneous polynomial of degree $i$ in $f_x,f_t,\cdots$, WITHOUT $f$.

Using the logarithmic transformation $u=2\ln{f}$,  the following expressions are obtained:
\begin{align}
&u_{xx}=2\left({\dfrac{f_{xx}}{f}}-\dfrac{f_x^2}{f^2}\right) =
\dfrac{D_x^2f\cdot
f}{f^2} \label{uxx}\\
&u_{xt}=2\left({\dfrac{f_{xt}}{f}}-\dfrac{f_xf_t}{f^2}\right)
=
\dfrac{D_xD_tf\cdot f}{f^2}  \\
&u_{tt}=2\left({\dfrac{f_{tt}}{f}}-\dfrac{f_t^2}{f^2}\right) =
\dfrac{D_t^2f\cdot f}{f^2}  \\
&u_{4x}=2\left({\dfrac{f_{xxxx}}{f}}-\dfrac{4f_{xxx}f_x+3f_{x,x}^2}{f^2}+\dfrac{12f_{xx}f_{x}^2}{f^3}-\dfrac{6f_x^4}{f^4}\right)
=\dfrac{D_x^4f\cdot f}{f^2}-3\left(  \dfrac{D_x^2f\cdot f}{f^2}
\right)^2\\
&u_{6x}=2\left({\dfrac{f_{xxxxxx}}{f}}+\cdots-\dfrac{120f_x^6}{f^6}\right)=\dfrac{D_x^6f\cdot
f}{f^2}-15\dfrac{D_x^4f\cdot f}{f^2}\dfrac{D_x^2f\cdot f}{f^2}
+30\left(  \dfrac{D_x^2f\cdot f}{f^2} \right)^3  \label{u6x}.
\end{align}

Clearly, the expressions above can lead us to finding the KdV-type bilinear form associated
with logarithmic transformation in NPDE.

\subsection{The Relationship Between Logarithmic Transformation and KdV-type Bilinear
Equation} 

In this paper, we consider the logarithmic transformation

\begin{equation}\label{logtrans}
u=2\alpha\left(\ln{f}\right)_{nx}, \qquad n=0,1,2
\end{equation}
where $f=f(x,t,\cdots)$ and $\alpha$ is a nonzero constant.

Substituting (\ref{logtrans}) to a NPDE
\begin{equation}\label{w}
W(u,u_{x},u_{t},\cdots)=0,
\end{equation}
we get
\begin{equation}\label{wfun}
\widetilde{W}(f,f_{x},f_{t},\cdots;\alpha) \equiv W(u,u_{x},u_{t},\cdots)|_{u=2\alpha\left(\ln{f}\right)_{nx}}.
\end{equation}

If NPDE $\widetilde{W}(f,f_{x},f_{t},\cdots;\alpha)$ has the KdV-type
bilinear form

\begin{equation}\label{KdV-type}
\left(\dfrac{P(\overrightarrow{D})f \cdot f}{f^2}\right)_{mx}=0,
\end{equation}
where $m$ is a nonnegative integer, using the properties of Hirota operator and logarithmic transformation, our algorithm will find the
undetermined parameters $n,m,P(\overrightarrow{D}), \mbox{and } \alpha$.

Because of (\ref{lnf_diff2}), the terms in NPDE with degree  $k$ generate
the homogeneous expression with logarithmic transformation
(\ref{logtrans})
\begin{equation}\label{lnf_kx}
\dfrac{\widetilde{P}_k}{f^k}+\dfrac{\widetilde{P}_{k+1}}{f^{k+1}}+\cdots.
\end{equation}
That is,with logarithmic transformation (\ref{logtrans}), the terms in NPDE with degree $k$ do NOT generate the terms
$\frac{\widetilde{P}_i}{f^i}$ $(i=1,2,3,\cdots,k-1)$.

Consider KdV-type bilinear form
\begin{equation}\label{HBF-expand}
\begin{split}
&\left(\dfrac{P(\overrightarrow{D})f \cdot f}{f^2}\right)_{mx}
=\left(\dfrac{P_1f+P_2}{f^2}\right)_{mx}
=\left(\dfrac{P_1}{f}+\dfrac{P_2}{f^2}\right)_{mx}=\left(\dfrac{P_1}{f}\right)_{mx}+\left(\dfrac{P_2}{f^2}\right)_{mx}\\
=&
\left(\dfrac{{P_1}_{mx}}{f}+\cdots+\dfrac{(-1)^{m}m!{P_1}f_x^m}{f^{m+1}}\right)+
\left(\dfrac{{P_2}_{mx}}{f^2}+\cdots+\dfrac{(-1)^{m}(m+1)!P_2f_x^m}{f^{m+2}}\right)\\
=&
\dfrac{{P_1}_{mx}}{f}+\cdots+\dfrac{(-1)^{m}(m+1)!P_2f_x^m}{f^{m+2}}=\dfrac{2\partial_x^m({P(\overrightarrow{\partial})f})}{f}+\cdots+\frac{(-1)^{m}(m+1)!P_2f_x^m}{f^{m+2}}.
\end{split}
\end{equation}

Therefore, if a equation has the KdV-type bilinear form, with logarithmic transformation (\ref{logtrans}), the term
$\frac{{P_1}_{mx}}{f}$ must be generated from the linear terms in NPDE (\ref{w}).


Denoting the linear part in  NPDE (\ref{w}) by $P_l(\overrightarrow{\partial})u$,
with logarithmic transformation (\ref{logtrans}), according to (\ref{lnf_diff2}), the linear part is
transformed into
\begin{equation*}
P_l(\overrightarrow{\partial})u \xrightarrow{u=2\alpha\left(\ln{f}\right)_{nx}}
\frac{2\alpha\partial_x^{n}P_l(\overrightarrow{\partial})f}{f}+\cdots.
\end{equation*}
Thus, it follows that
\begin{equation}\label{P_P_l}
P(\overrightarrow{\partial})=\alpha\partial_x^{n-m}P_l(\overrightarrow{\partial}).
\end{equation}

From the above equation, we find it important that the linear terms of the NPDE play a crucial role in finding the HBF. Hence, from the linear part in NPDE, $P$, the function of HBF, can be obtained.

Let
\begin{equation}
\left(\dfrac{P(\mathbf{D})f \cdot f}{f^2}\right)_{mx}
=\widetilde{W}(f,f_{x},f_{t},\cdots;\alpha),
\end{equation}
then the difference should be zero, namely,
\begin{equation}\label{Residual}
Res_{\widetilde{W}}  \triangleq
\widetilde{W}(f,f_{x},f_{t},\cdots;\alpha) -
\left(\dfrac{P(\overrightarrow{D})f \cdot f}{f^2}\right)_{mx}=0.
\end{equation}

Clearly, $n$ is a finite enumerated integer and $m$ is also a finite enumerated integer (See \textbf{Step 5} of algorithm in \textbf{Section 5}) and $P(\overrightarrow{\partial})=\alpha\partial_x^{n-m}P_l(\overrightarrow{\partial})$.

With $P$ satisfying (\ref{P_P_l}), the coefficient of  $\frac{1}{f}$ in $Res_{\widetilde{W}}$ is equal to zero. Then the
homogeneous expression in $f$ and its derivatives follows as:

\begin{equation}\label{Residual2}
Res_{\widetilde{W}}=\dfrac{\widehat{P}_2}{f^2}+\cdots+\dfrac{\widehat{P}_r}{f^r}.
\end{equation}

Here, $r$ is a finite positive integer. 

Solving the numerators of (\ref{Residual2}), if there exists the nonzero numerical solution $\alpha$, the
undetermined parameters $n,m,P(\overrightarrow{D})$ are also achieved. It can be proven that  $\alpha$ has no more than one nonzero numerical solution which satisfies (\ref{Residual}) with certain $n$ and
$m$.The proof will be given in our next paper.

In summary, if a NPDE has the KdV-type bilinear form associated with the logarithmic transformation (\ref{logtrans}), our algorithm will surely find it.

\section{An Example Of our Algorithm}
\label{}

Now, let us take the KdV equation as an example.

\begin{equation}\label{KdVeqn}
u_{xxx}+6uu_{x}+u_{t}=0.
\end{equation}
Here, $P_{l}(\overrightarrow{\partial})=\partial_{t}+\partial_{x}^3$.

First substitute $u=2\alpha\ln{f}$
to (\ref{KdVeqn}) and simplify the equation, the
homogeneous equation in $f$ and its derivatives follows as:


\[\frac{2\alpha(f_{xxx}+12\alpha(\ln{f})f_{x}+f_{t})}{f}-\frac{6\alpha
f_{xx}f_{x}}{f^2}+\frac{4\alpha f_{x}^3}{f^3}=0.\]
Since the above equation has $\ln{(f)}$ term, we goto the next
logarithmic transformation $u=2\alpha(\ln{f})_{x}$.

Likewise,substitute $u=2\alpha(\ln{f})_{x}$ to (\ref{KdVeqn}) and simplify the equation, the
homogeneous equation in $f$ and its derivatives follows as:
\[\frac{2\alpha(f_{xxxx}+f_{tx})}{f}-\frac{2\alpha(-12\alpha
f_{xx}f_{x}+4f_{xxx}f_{x}+3f_{xx}^2+f_{x}f_{t})}{f^2}+\frac{24\alpha
f_{x}^2(-\alpha f_{x}+f_{xx})}{f^3}-\frac{12\alpha f_{x}^4}{f^4}=0.
\]
From the above equation, the coefficient of $\dfrac{1}{f}$ is acquired:
$P_1=2\alpha(f_{xxxx}+f_{tx})=2\alpha\left(\partial_x^4+\partial_x\partial_t\right)f=2\alpha\partial_{x}P_{l}(\overrightarrow{\partial})f$. Because
the order of derivative is even, the Hirota bilinear form can be obtained:
$P(\overrightarrow{D})f \cdot f=\alpha(D_{x}^4+D_{x}D_{t})f \cdot f$, with which divided
by $f^2$ and subtracts the above equation  the
difference is as follows: $$\frac{12f_{xx}\alpha(-2\alpha
f_{x}+f_{xx})}{f^2}-\frac{24\alpha f_{x}^2(-\alpha
f_{x}+f_{xx})}{f^3}+\frac{12\alpha f_{x}^4}{f^4}=0.$$
Solving the numerators of the difference equation, it only has
$\alpha=0$. So we goto the next logarithmic transformation
$u=2\alpha(\ln{f})_{xx}$.

Similarly, substitute  $u=2\alpha(\ln{f})_{xx}$ to (\ref{KdVeqn}) and simplify the equation,the
homogeneous equation in $f$ and its derivatives follows as:
$$ \frac{2\alpha(f_{txx}+f_{xxxxx})}{f}-\frac{2\alpha(-12\alpha
f_{xxx}f_{xx}+f_{xx}f_{t}+5f_{xxxx}f_{x}+10f_{xxx}f_{xx})+2f_{x}f_{tx}}{f^2}$$
$$-\frac{4f_{x}\alpha(18\alpha f_{xx}^2+6\alpha
f_{xxx}f_{x}-15f_{xx}^2-10f_{xxx}f_{x}-f_{x}f_{t})}{f^3}+\frac{120\alpha
f_{xx}f_{x}^3(\alpha-1)}{f^4}-\frac{48\alpha
f_{x}^5(\alpha-1)}{f^5}=0.$$ From the above equation, the coefficient of
$\dfrac{1}{f}$ is obtained:
$P_1=2\alpha(f_{xxxxx}+f_{txx})=2\alpha\left(\partial_x^5+\partial_x^2\partial_t\right)f$. Because
the order of derivative is odd, we integrate $P_1$ and get the
Hirota bilinear form $P(\overrightarrow{D})f \cdot
f=\alpha(D_{x}^4+D_{x}D_{t})f \cdot f$, with which divided by $f^2$ and derived
in terms of $x$ one time then subtracts the above equation,
the difference is obtained:
$$\frac{-24\alpha
f_{xxx}f_{xx}(\alpha-1)}{f^2}+\frac{24f_{x}\alpha(\alpha-1)(3f_{xx}^2+f_{xxx}f_{x})}{f^3}-\frac{120\alpha
f_{xx}f_{x}^3(\alpha-1)}{f^4}+\frac{48\alpha
f_{x}^5(\alpha-1)}{f^5}=0.$$
Solving the numerators of the difference equation, a nonzero
solution, $\alpha=1$, is achieved.

In conclusion, by using the logarithmic transformation $u=2(\ln{f})_{xx}$, it shows that the
equation (\ref{KdVeqn}) is equivalent to
 $$\left(\frac{(D_{x}^4+D_{x}D_{t})f\cdot
f}{f^2}\right)_{x}=0.$$

This example illustrates our method intuitively.

\section{An Algorithm of Bilinear}
\label{}

\renewcommand{\labelenumi}{\textbf{Step \arabic{enumi} }}
\renewcommand{\labelenumii}{\textbf{Step \arabic{enumi}.\arabic{enumii} }}

Consider the general NPDE
$$W(u,u_x,u_t,\cdots)=0,$$
where $u=u(x,t,\cdots)$, and $W$ is a polynomial in $u$ and its
derivatives.

The algorithm for obtaining HBF  goes in the following steps:

\begin{enumerate}
\item
Let $n=0$, $f=f(x,t,\cdots)$ and $\alpha$ is an undetermined nonzero
constant.

\item
Using logarithmic transformation $u = 2\alpha
\left(\ln{f}\right)_{nx}$, it follows that
$$\widetilde{W}=\widetilde{W}(f,f_x,f_t,\cdots)=W(u,u_x,u_t,\cdots)|_{u=2\alpha(\ln{f})_{nx}}$$

\item
If $\widetilde{W}$ has  $\ln{(f)}$ term, then goto \textbf{Step 6}.

\item
Simplify the equation $\widetilde{W}$ to get the homogeneous expressions
in $f$ and its derivatives

$$\widetilde{W}=\dfrac{P_1}{f}+\dfrac{P_2}{f^2}+\cdots+\dfrac{P_r}{f^r},$$
where $P_i=P_i\left(f_x,f_t,\cdots\right)$ $(i=1,2,\cdots,r)$ is a homogeneous polynomial of degree $i$ in $f_x,f_t,\cdots$, WITHOUT $f$.

\item
Get the coefficient of $\dfrac{1}{f}$ in $\widetilde{W}$ :
$P_1=P_1(f_x,f_t,\cdots)$. Let the lowest order of the derivative of $f$ with respect to $x$ in $P_{1}$
 is $k$, and the corresponding  Hirota bilinear form is
$P(D_x,D_t,\cdots)f \cdot f$. Then the following equation is obtained:
$$P_1(f_x,f_t,\cdots)=2\partial_{x}^{m}P(f_x,f_t,\cdots),$$
where $0 \leq m \leq k$ and the degrees of the terms in $P(f_x,f_t,\cdots)$ are
even. That is,
\begin{itemize}
\item
If the lowest order derivative of $f$ is even, then $m=0,2,4,\cdots$,
and $m \leq k$.
\item
If the lowest order  derivative of $f$ is odd, \:  then
$m=1,3,5,\cdots$, and $m \leq k$.
\end{itemize}

\begin{enumerate}
\item
Let $m=0$, if the  lowest order of the derivative of $f$ is even, or else let
$m=1$.
\item
Integrating $P_1$ by $m$ times in terms of $x$, we get
$P(f_x,f_t,\cdots)=\frac{1}{2}\partial_x^{-m}P_1(f_x,f_t,\cdots)$. Hence the HBF
is $P(D_x,D_t,\cdots)f \cdot f$.

\item
Calculate the difference, $Res_{\widetilde{W}} = \widetilde{W} -
\left(\dfrac{P(D_x,D_t,\cdots)f \cdot f}{f^2}\right)_{mx}$, and
simplify the difference  $Res_{\widetilde{W}}$ to get the  homogeneous
expressions in $f$ and its derivatives
$$Res_{\widetilde{W}}=\dfrac{\widehat{P}_2}{f^2}+\cdots+\dfrac{\widehat{P}_r}{f^r}.$$

\item
Get the equations $\widehat{P}_i=0$ $(i=2,3,\cdots,r)$, and solve the
equations. If a nonzero  $\alpha$ exists, then using
$u=2\alpha\left(\ln{f}\right)_{nx}$, the NPDE is equal to
$\left(\dfrac{P(D_x,D_t,\cdots)f \cdot f}{f^2}\right)_{mx}=0$. Therefore,
the HBF is $P(D_x,D_t,\cdots)f \cdot f$, and exit of program.
\item
Let $m=m+2$. If $m \leq k$, goto \textbf{Step 5.2}, or else goto
\textbf{Step 6}.
\end{enumerate}

\item
Let $n=n+1$. If $n \le 2$ then goto \textbf{Step 2}, or else the NPDE
has no HBF with logarithmic transformation, end program and exit.
\end{enumerate}

\section{Applications}
\label{}
\theoremstyle{definition}
\newtheorem{example}{Example}
\newtheorem*{prf}{Proof}
\newtheorem*{solution}{Solution}

\begin{example}
Boussinesq  equation \cite{13,14}

\begin{equation}
u_{tt}-u_{xx}-3u_{xx}^2-u_{4x}=0
\end{equation}

\end{example}

\begin{solution}
From the equation, $Bilinear$ within 0.s outputs $\alpha=1 $,
$u=2\ln{(f)} $ and
$$\frac{(D_{t}^2-D_{x}^2-D_{x}^{4})f\cdot f}{f^2}=0.$$

\end{solution}

\begin{example}
Sawada-Kotera equation \cite{15,16}
\begin{equation}
u_{t}+45u_{x}u^2-15u_{xx}u_{x}-15u_{xxx}u+u_{5x}=0
\end{equation}
\end{example}
\begin{solution}
Within 0.032s our program outputs $\alpha=-1 $, $u=-2(\ln{f})_{xx}$
and
$$\left(\frac{(-D_{x}^6-D_{x}D_{t})f\cdot f}{f^2}\right)_{x}=0.$$
\end{solution}

\begin{example}
Kadomtsev-Petviashvili equation \cite{17}
\begin{equation}
(u_{3x}+6uu_{x}+u_{t})_{x}+3\delta^{2} u_{yy}=0
\end{equation}
\end{example}
\begin{solution}
$Bilinear$ within 0.016s outputs $\alpha=1 $, $u=2(\ln{f})_{xx} $
and
$$\left(\frac{(D_{x}^4+D_{x}D_{t}+3\delta^{2}D_{y}^2)f\cdot f}{f^2}\right)_{xx}=0.$$

\end{solution}

\begin{example}The shallow water waves equation \cite{18}
\begin{equation}\label{exmaple1}
u_{t}-u_{xxt}-3uu_{t}+3u_{x}\int_{x}^{\infty}u_{t}dx+u_{x}=0
\end{equation}
\end{example}
\begin{solution}
By substituting $u=w_{x}$ and using the boundary condition
$u_{t}|_{x\rightarrow \infty}=0$ (\ref{exmaple1}) can be converted into the
differential form
\begin{equation}
w_{xt}-w_{3xt}-3w_{x}w_{xt}-3w_{xx}w_{t}+w_{xx}=0,
\end{equation}
$Bilinear$ within 0.016s outputs $\alpha=1 $, $w=2(\ln{f})_{x} $ and
$$\left(\frac{(-D_{x}^3D_{t}+D_{x}^2+D_{x}D_{t})f\cdot
f}{f^2}\right)_{x}=0.$$

\end{solution}

\begin{example}
Ito equation \cite{19}
\begin{equation}\label{exmaple2}
u_{tt}+u_{3xt}+6u_{x}u_{t}+3uu_{xt}+3u_{xx}\int_{-\infty}^{x}u_{t}dx=0
\end{equation}
\end{example}
\begin{solution}
By substituting $u=w_{x}$ and using the boundary condition
$u_{t}|_{x\rightarrow -\infty}=0$ (\ref{exmaple2}) is transformed into the
differential form
\begin{equation}
w_{xtt}+w_{4xt}+6w_{xx}w_{xt}+3w_{x}w_{xxt}+3w_{3x}w_{t}=0,
\end{equation}
$Bilinear$ within  0.031s  outputs $\alpha=1 $, $w=2(\ln{f})_{x} $
and
$$\left(\frac{(D_{x}^3D_{t}+D_{t}^2)f\cdot
f}{f^2}\right)_{xx}=0.$$

\end{solution}

\begin{example}
(2+1)-dimensional breaking soliton equation \cite{20}
\begin{equation}
u_{t}+\beta u_{xxy}+4\beta uu_{y}+4\beta
u_{x}\int_{-\infty}^{x}u_{y}dx=0
\end{equation}
\end{example}
\begin{solution}
The equation can be written as (6.9) by substituting $u=w_{x}$ and using the boundary condition
$u_{y}|_{x\rightarrow -\infty}=0$
\begin{equation}
w_{xt}+\beta w_{3xy}+4\beta w_{x}w_{xy}+4\beta w_{xx}w_{y}=0,
\end{equation}
$Bilinear$ within 0.016s outputs $\alpha=\frac{3}{4} $,
$w=\frac{3}{2}(\ln{f})_{x} $ and
$$\left(\frac{3}{4}\frac{(D_{x}D_{t}+\beta D_{x}^3D_{y})f\cdot
f}{f^2}\right)_{x}=0.$$

\end{solution}

\begin{example}
The Bidirectional SK equation \cite{21}
\begin{equation}\label{exmaple3}
5\int_{-\infty}^{x}u_{tt}dx+5u_{xxt}-15uu_{t}-15u_{x}\int_{-\infty}^{x}u_{t}dx-45u_{x}u^2+15u_{xx}u_{x}+15u_{3x}u-u_{5x}=0
\end{equation}
\end{example}
\begin{solution}
Similarly, we  convert (\ref{exmaple3}) into the differential form by
substituting $u=w_{x}$ and using the boundary condition
$u_{t}|_{x\rightarrow -\infty}=0,u_{tt}|_{x\rightarrow -\infty}=0$
\begin{equation}
5w_{tt}+5w_{3xt}-15w_{x}w_{xt}-15w_{xx}w_{t}-45w_{xx}w_{x}^2+15w_{3x}w_{xx}+15w_{4x}w_{x}-w_{6x}=0,
\end{equation}
$Bilinear$ within 0.015s outputs $\alpha=-1 $, $w=-2(\ln{f})_{x} $
and
$$\left(\frac{(-5D_{t}^2+D_{x}^6-5D_{x}^3D_{t})f\cdot
f}{f^2}\right)_{x}=0.$$
\end{solution}

\begin{example}
(2+1)-dimensional  SK equation \cite{22}
\begin{equation}\label{exmaple4}
9u_{t}+u_{5x}+15u_{xx}u_{x}+15u_{3x}u+45u_{x}u^2-5\int_{-\infty}^{x}u_{yy}dx-15uu_{y}-15u_{x}\int_{-\infty}^{x}u_{y}dx-5u_{xxy}=0
\end{equation}
\end{example}
\begin{solution}
With the substitution $u=w_{x}$ and the boundary condition
$u_{y}|_{x\rightarrow -\infty}=0,u_{yy}|_{x\rightarrow -\infty}=0$
 (\ref{exmaple4}) acquires a differential form
\begin{equation}
9w_{xt}+w_{6x}+15w_{3x}w_{xx}+15w_{4x}w_{x}+45w_{xx}w_{x}^2-5w_{yy}-15w_{x}w_{xy}-15w_{xx}w_{y}-5w_{3xy}=0,
\end{equation}
$Bilinear$ within 0.032s outputs $\alpha=1 $, $w=2(\ln{f})_{x} $ and
$$\left(\frac{(D_{x}^6-5D_{x}^3D_{y}-5D_{y}^2+9D_{x}D_{t})f\cdot
f}{f^2}\right)_{x}=0.$$

\end{solution}

\begin{example}
(3+1)-dimensional KdV equation \cite{9}
\begin{equation}
u_{t}+6u_{x}u_{y}+u_{xxy}+u_{4xz}+60u_{x}^2u_{z}+10u_{3x}u_{z}+20u_{x}u_{xxz}=0
\end{equation}
\end{example}
\begin{solution}
$Bilinear$ within 0.016s outputs $\alpha=\frac{1}{2} $,
$u=(\ln{f})_{x}$ and
$$\frac{1}{2}\frac{(D_{x}D_{t}+D_{x}^5D_{z}+D_{x}^3D_{y})f\cdot
f}{f^2}=0.$$

\end{solution}

\section{The Program Code of Bilinear}
\label{}


\begin{verbatim}
with(PDEtools): with(DEtools):

## Hirota Bilinear Method
## Bilinear Derivative / Hirota Operator

BD:=proc(FF,DD) local f,g,x,m,opt;
if nargs=1 then return `*`(FF[]); fi; f,g:=FF[]; x,m:=DD[];
opt:=args[3..-1]; if m=0 then return procname(FF,opt); fi;
procname([diff(f,x),g],[x,m-1],opt)-procname([f,diff(g,x)],[x,m-1],opt);
end:

`print/BD`:=proc(FF,DD) local f,g,x,m,i; f,g:=FF[];
f:=cat(f,` ・ `,g); g:=product(D[args[i][1]]^args[i][2],i=2..nargs);
if g<>1 then f:=``(g)*``(f); fi; f; end:

## collect(expr,f); first!
getFnumer:=proc(df,f,pow::posint:=1) local i,g,fdenom;
if type(df,`+`) then
 g:=[op(df)];
 fdenom:=map(denom,g);
 for i to nops(fdenom) while fdenom[i]<>f^pow do od;
 if i>nops(fdenom) then lprint(fdenom);
 error "no term(s) or numer=0 when denom=%1",op(0,f)^pow fi;
 g:=numer(g[i]);
 if not type(expand(g),`+`) then  lprint(g);
 error "Expected more than 1 term about Hirota D-operator" fi;
 return g; fi; lprint(df);
 error "expected 1st argument be type `+`."; end:

getvarpow:=proc(df::function) local i,f,var,dif,pow; if
op(0,df)<>diff then lprint(df); error "expected diff function" fi;
f:=convert(df,D); var:=[op(f)]; dif:=[op(op([0,0],f))];
pow:=[0$nops(var)]; f:=op(op(0,f))(var[]); for i to nops(var) do
  dif:=selectremove(member,dif,{i});
  pow[i]:=nops(dif[1]);
  dif:=dif[2];
  od; pow:=zip((x,y)->[x,y],var,pow); pow:=remove(has,pow,{0});
  [[f,f],pow[]]; end:

#convert to Hirota Bilinear Form
HBF:=proc(df) local i,c,f; if type(df,`+`) then
  f:=[op(df)];  return map(procname,f); fi;
  if type(df,`*`) then f:=[op(df)];
  f:=selectremove(hasfun,f,diff); c:=f[2]; f:=f[1];
  if nops(f)<>1 then lprint(df); error "need only one diff function factor." fi;
  f:=f[]; c:=`*`(c[]); f:=getvarpow(f); f:=[c,f]; return f; fi;
  if op(0,df)=diff then f:=getvarpow(df); f:=[1,f]; return f; fi;
  lprint(df); error "unexpected type."; end:

printHBF:=proc(PL::list) local j,DD,f,C,tmp,gcdC; C:=map2(op,1,PL);
gcdC:=1; if nops(C)>1 then tmp:=[seq(cat(_Z,i),i=1..nops(C))];
  gcdC:=tmp *~ C; gcdC:=`+`(gcdC[]); gcdC:=factor(gcdC);
  tmp:=selectremove(has,gcdC,tmp); gcdC:=tmp[2];
  if gcdC=0 then gcdC:=1 fi; gcdC:=gcdC*content(tmp[1]); fi;
  if gcdC<>1 then  C:=C /~ gcdC; fi; DD:=map2(op,2,PL);
f:=op(0,DD[1][1][1]);
DD:=map(z->product(D[z[i][1]]^z[i][2],i=2..nops(z)),DD);
DD:=zip(`*`,C,DD); DD:=`+`(DD[]); gcdC * ``(DD) * cat(f,` ・ `,f);
end:

## print Hirota Bilinear Transform
printHBT:=proc(uf,u,f,i,j,PL,alpha:=1) local DD,g,C,tmp,pl;
pl:=printHBF(PL); if j>0 then print(u=2*alpha*'diff'(ln(f),x$j));
else print(u=2*alpha*ln(f)); fi;
if i>0 then print('diff'(pl/f^2,x$i)=0); else
print(pl/f^2=0); fi; NULL; end:

guessdifforder:=proc(PL::list,x::name)
local L,minorder,maxorder,tmp; L:=map2(op,2,PL);
L:=map(z->z[2..-1],L); tmp:=map(z->map2(op,2,z),L);
tmp:=map(z->`+`(z[]),tmp); tmp:=selectremove(type,tmp,even);
minorder:=0; if nops(tmp[1])<nops(tmp[2]) then minorder:=1 fi;
tmp:=map(z->select(has,z,{x}),L); tmp:=map(z->map2(op,2,z),tmp); if
has(tmp,{[]}) then maxorder:=0; else tmp:=map(op,tmp);
  maxorder:=min(tmp[]); fi;
if type(maxorder-minorder,odd) then maxorder:=maxorder-1 fi;
[minorder,maxorder]; end:

guessalpha:=proc(Res,uf,u,f,i,j,PL,alpha) local tmp,res,pl,flag,k;
flag:=1; tmp:=[op(Res)]; tmp:=map(numer,tmp);
tmp:=gcd(tmp[1],tmp[-1]); if type(tmp,`*`) then
tmp:=remove(has,tmp,f); fi; if tmp<>0 and has(tmp,{alpha}) then
  tmp:=solve(tmp/alpha^difforder(uf),{alpha});
  if tmp<>NULL and has(tmp,{alpha}) then lprint(tmp);
    for k to nops([tmp]) while flag=1 do
      res:=collect(expand(subs(tmp[k],Res)),f,factor);
      if res=0 then pl:=subs(tmp[k],PL);
        printHBT(uf,u,f,i,j,pl,rhs(tmp[k]));
        flag:=0; fi; od; fi; fi; PL; end:

Bilinear:=proc(uf,u,f,x,alpha) local su,h,i,j,g1,CB,PL,gdo,DD,Res;
if hasfun(uf,int) then error "Do not support integral function yet.
Please substitute int function." fi; for j from 0 to 2 do
  Res:=1; su:=u=2*alpha*diff(ln(f),[x$j]);
  h:=collect(expand(dsubs(su,uf)),f,factor);
  if hasfun(h,ln) then next; fi;
  g1:=getFnumer(h,f)/2; g1:=expand(g1); CB:=HBF(g1);
  gdo:=guessdifforder(CB,x);
  for i from gdo[1] by 2 to gdo[2] do
    if i=0 then PL:=CB; else PL:=HBF(int(g1,x$i)); fi;
    DD:=add(PL[i][1]*BD(PL[i][2][]),i=1..nops(PL));
    Res:=collect(expand(diff(DD/f^2,[x$i])-h),f,factor);
    if Res=0 then printHBT(uf,u,f,i,j,PL,alpha); break;
    elif type(alpha,name) and has(DD,alpha) then
      Res:=guessalpha(Res,uf,u,f,i,j,PL,alpha);
    fi; od; if Res=0 then break; fi; od; PL; end:
\end{verbatim}

\section{Conclusions}
\label{}

 To sum up, an algorithm for generating the
Hirota bilinear form of NPDE with logarithm transformation has been
proposed in this paper, and the bilinear forms of a class of NPDEs are
obtained by using package Bilinear. Then we illuminate the availability
of the algorithm by illustrating some examples.

\section*{Acknowledgement}

 The work has been partially supported by the Natural Science Foundation
of China ( No. 10971109),  K.C.Wong Magna Fund
in Ningbo University and Student Research of Ningbo University.
Jingsong He is also supported by Program for NCET under Grant No.NCET-08-0515.




\begin{thebibliography}{00}


\bibitem{1}R. Hirota, The direct method in soliton theory (Cambridge University Press, Cambridge, 2004).
\bibitem{}A. Scott, Encyclopedia of Nonlinear Science, Taylor and Francis, Routledge, New York, 2005.
\bibitem{}Wenxiu Ma, Ruguang Zhou, Liang Gao, Exact one-periodic and two-periodic wave solutions to Hirota bilinear equations in (2+1) dimensions, Mod.
          Phys. Lett. A 21 (2009) 1677-1688.
\bibitem{}Wenxiu Ma, Engui Fan, Linear superposition principle applying to Hirota bilinear equations, Comput. Math. Appl. 61 (2011) 950-959.
\bibitem{}J. Hietarinta, Introduction to the Hirota bilinear method. Integrability of nonliear systems (Pondicherry,1996) 95-103, Lecture Notes in Physics, 495 (Springer, Berlin, 1997).
\bibitem{2}J. Hietarinta, A search for bilinear equations passing Hirota's three-soliton condition. I. KdV-type bilinear equations. J. Math. Phys. 28 (1987) 1732-1742.
\bibitem{3}J. Hietarinta, A search for bilinear equations passing Hirota's three-soliton condition. II. mKdV-type bilinear equations. J. Math. Phys. 28 (1987) 2094-2101.
\bibitem{4}J. Hietarinta, A search for bilinear equations passing Hirota's three-soliton condition. III. Sine-Gordon-type bilinear equations. J. Math. Phys. 28 (1987) 2586-2592.
\bibitem{5}J. Hietarinta, A search for bilinear equations passing Hirota's three-soliton condition. IV. Complex bilinear equations. J. Math. Phys. 29 (1988) 628-635.
\bibitem{6}W. Hereman and W. Zhuang, Symbolic computation of solitons with Macsyma. Computational and applied mathematics, II (Dublin, 1991), 287-296 (North-Holland, Amsterdam, 1992).
\bibitem{7}Zhenjiang Zhou, Jingzhi Fu and Zhibin Li, An implementation for the algorithm of Hirota bilinear form of PDE in the Maple system. Appl. Math. Comput. 183 (2006) 872-877.
\bibitem{8}Zhenjiang Zhou, Jingzhi Fu and Zhibin Li, Maple packages for computing Hirota's bilinear equation and multisoliton solutions of
nonlinear evolution equations. Appl. Math. Comput. (2010) 92-104.
\bibitem{9}Xudong Yang and Hangyu Ruan, A Maple package on symbolic computation of Hirota bilinear form for nonlinear
equations. Commun. Theor. Phys. (Beijing, China) 52 (2009) 801-807.
\bibitem{10}J. Weiss, M. Tabor and G. Carnevale, The Painleve property for partial differential equations. J. Math. Phys. 24 (1983) 522-526.
\bibitem{11}Mingliang Wang, Yubin Zhou and Zhibin Li, Application of a homogeneous balance method to exact solutions of nonlinear equations in mathematical physics. Phys. Lett. A 216 (1996) 67-75.
\bibitem{13}G. B. Whitham, Linear and Nonlinear Waves (New York, Wiley, 1974) 9.
\bibitem{14}D. Zwillinger, Handbook of Differential Equations, 3rd ed. (Boston, MA: Academic Press,1997) 129-130.
\bibitem{15}K. Sawada and T. Kotera, A method for finding $N$-soliton solutions of the K.d.V. equation and K.d.V.-like equation. Progr. Theoret. Phys. 51 (1974) 1355-1367.
\bibitem{16}P. J. Caudrey, R. K. Dodd and J. D. Gibbon, A new hierarchy of Korteweg-de Vries equations. Proc. Roy. Soc. London Ser. A 351 (1976) 407-422.
\bibitem{17}B. B. Kadomtsev and V. I. Petviashvili, On the stability of solitary waves in weakly dispersive media. Sov. Phys. Dokl. 15 (1970) 539-541.
\bibitem{18}R. Hirota and J. Satsuma, $N$-soliton solutions of model equations for shallow water waves. J. Phys. Soc. Japan. 40 (1976) 611-612.
\bibitem{19}M. Ito, An extension of nonlinear evolution equations of the K-dV (mK-dV) type to higher orders. J. Phys. Soc. Japan. 49 (1980) 771-778.
\bibitem{20}WenHua Huang, YuLu Liu and JieFang Zhang, Doubly periodic propagating wave for $(2+1)$-dimensional breaking soliton equation. Commun. Theor. Phys. (Beijing) 49 (2008) 268-274.
\bibitem{21}J. M. Dye and A. Parker, A bidirectional Kaup-Kupershmidt equation and directionally dependent solitons. J. Math. Phys. 43 (2002) 4921-4949.
\bibitem{22}Jingsong He and Xiaodong Li, Solutions of the $(2+1)$-dimensional KP, SK and KK equations generated by gauge transformations from nonzero seeds. J. Nonlinear Math. Phys. 16 (2009) 179-194.


\end{thebibliography}
\end{document}